\documentclass[twocolumn,showpacs,preprintnumbers,amsmath,amssymb,prd]{revtex4}
\usepackage{amsmath}\usepackage{amssymb}\usepackage{amsfonts}
\usepackage{natbib}\usepackage{graphics}\usepackage{latexsym}\usepackage{dcolumn}\usepackage{bm}
\usepackage[dvips]{epsfig}

\begin{document}
\setlength{\unitlength}{1mm}


\newcommand{\ket}[1] {\mbox{$ \vert #1 \rangle $}}
\newcommand{\bra}[1] {\mbox{$ \langle #1 \vert $}}
\def\vac{\ket{0}} \def\vacin{\ket{0}_{in}} \def\vacout{\ket{0}_{out}}
\def\thermal{\ket{\beta}}
\def\bvac{\bra{0}}\def\bvacin{{}_{in}\bra{0}}\def\bvacout{{}_{out}\bra{0}}
\def\bthermal{\bra{\beta}}
\newcommand{\ave}[1] {\mbox{$ \langle #1 \rangle $}}
\newcommand{\avew}[1] {\mbox{$ \langle #1 \rangle $}_w}
\newcommand{\vacave}[1] {\mbox{$ \bvac #1 \vac $}}
\newcommand{\thermalave}[1] {\mbox{$ \bthermal #1 \thermal $}}
\newcommand{\scal}[2]{\mbox{$ \langle #1 \vert #2 \rangle $}}
\newcommand{\expect}[3] {\mbox{$ \bra{#1} #2 \ket{#3} $}}
\def\a{\hat{a}}\def\A{\hat{A}}
\def\b{\hat{b}}\def\B{\hat{B}}
\def\aa{\tilde{a}}\def\AA{\tilde{A}}
\def\bb{\tilde{b}}\def\BB{\tilde{B}}
\def\kplus{\ket{+}}\def\kmoins{\ket{-}}
\def\bplus{\bra{+}}\def\bmoins{\bra{-}}


\def\p{\prime}
\def\t{\tau}
\def\om{\omega}\def\Om{\Omega}
\def\ga{\gamma}
\def\omp{\om^\p}\def\Omp{\Om^\p}
\def\la{\lambda}\def\lap{\lambda^\p}
\def\mup{\mu^\p}\def\lp{l^\p}
\def\kp{k^\p}\def\sig{\sigma}
\def\ka{\kappa}
\def\al{\alpha}\def\alb{\bar\alpha}
\def\bt{\beta}\def\btb{\bar\beta}
\def\e{\epsilon}
\def\psip{\stackrel{.}{\psi}}\def\fp{\stackrel{.}{f}}
\def\lab{{\bar \la}}
\def\ffi{\varphi}
\def\scry{{\cal J}}
\def\scryp{{\cal J}^+}
\def\scrym{{\cal J}^-}
\def\scrypL{{\cal J}^+_L}
\def\scrymR{{\cal J}^-_R}
\def\scrypR{{\cal J}^+_R}
\def\scrymL{{\cal J}^-_L}
\def\Pig{{\mathbf \Pi}}

\def\di{\partial}
\def\diU{\di_U}
\def\diUU{\di_{\bar U}}
\def\diV{\di_V}
\def\diVV{\di_{\bar V}}
\def\didiv{\raise 0.1mm \hbox{$\stackrel{\leftrightarrow}{\di_V}$}}
\newcommand{\didi}[1]{\raise 0.1mm \hbox{$\stackrel{\leftrightarrow}{\di_{#1}}$}}

\def\d{\mbox{d}}


\def\TUU{T_{UU}}
\def\TUUI{T_{UU}^I}\def\TUUII{T_{UU}^{II}}
\def\Tuu{T_{uu}}
\def\TVV{T_{VV}}
\def\TVVI{T_{VV}^I}\def\TVVII{T_{VV}^{II}}
\def\Tvv{T_{vv}}
\def\re{\mbox{Re}}
\def\im{\mbox{Im}}
\def\S{{\mathbf S}}
\def\T{{\mathbf T}}
\def\1{{\mathbf 1}}
\def\Ss{\mbox{$\hat S$}}
\def\f{{\tilde f}}
\def\ubl{\bar u_L}
\def\UU{{\bar U}}
\def\VV{{\bar V}}
\def\LL{{\cal L_{\rm int}}}
\def\disp{\displaystyle}
\def\bitem{\begin{itemize}}
\def\eitem{\end{itemize}}
\def\bes{\begin{description}}
\def\es{\end{description}}
\newcommand{\be} {\begin{equation}}
\newcommand{\ee} {\end{equation}}
\newcommand{\ba} {\begin{eqnarray}}
\newcommand{\ea} {\end{eqnarray}}
\newcommand{\bsub}{\begin{subeqnarray}}
\newcommand{\esub}{\end{subeqnarray}}
\newcommand{\bwt} {\begin{widetext}}
\newcommand{\ewt} {\end{widetext}}

\def\cf{{\it cf}~}
\def\ie{{\it i.e.}~}
\def\etc{{\it etc}}
\def\eg{{\it e.g.}~}
\def\apriori{{\it a priori}~}

\def\nn{\nonumber \\}
\newcommand{\reff}[1]{Eq.(\ref{#1})}

\newcommand{\encadre}[1]{\begin{tabular}{|c|}\hline{#1}\\ \hline  \end{tabular}}

\def\sinc{\mbox{sinc}}
\def\sinhc{\mbox{sinhc}}
\def\arccosh{\mbox{arccosh}}
\def\arcsinh{\mbox{arcsinh}}
\def\arctanh{\mbox{arctanh}}
\def\l{\left}
\def\r{\right}
\def\Si{\mbox{Si}}
\def\half{{1 \over 2}}
\newcommand{\inv}[1]{\frac{1}{#1}}
\def\inte{\int_{-\infty}^{+\infty}}
\def\into{\int_{0}^{\infty}}
\newcommand\Ie[1]{\inte \! \mbox{d} #1 \;}
\newcommand\Io[1]{\into \! \mbox{d} #1 \;}
\newcommand\IeIe[2]{\int \!\!\! \inte \! \mbox{d} #1 \: \mbox{d} #2 \;}
\newcommand\IoIo[2]{\int \!\!\! \into \! \mbox{d} #1 \: \mbox{d} #2 \;}
\newcommand\Iomoins[1]{\int_{-\infty}^{0} \! \mbox{d} #1 \;}
\newcommand\Iinfsup[3]{\int_{#1}^{#2} \! \mbox{d} #3 \;}
\newcommand{\erf}{\mathop{\rm erf}\nolimits}

\overfullrule=0pt \def\sqr#1#2{{\vcenter{\vbox{\hrule height.#2pt
          \hbox{\vrule width.#2pt height#1pt \kern#1pt
           \vrule width.#2pt}
           \hrule height.#2pt}}}}
\newcommand\lrpartial[1]{\mathrel{\partial_{#1}\kern-1em\raise1.75ex\hbox{$\leftrightarrow$}}}

\newcommand{\doublesum}[4]{\sum_{\begin{array}{cc}\scriptstyle #2 \\ \scriptstyle #1
\end{array}}^{\begin{array}{cc}\scriptstyle #4 \\ \scriptstyle #3 \end{array}}}

\newcounter{subequation}[equation] \makeatletter
\expandafter\let\expandafter\reset@font\csname reset@font\endcsname
\newenvironment{subeqnarray}
  {\arraycolsep1pt
    \def\@eqnnum\stepcounter##1{\stepcounter{subequation}{\reset@font\rm
      (\theequation\alph{subequation})}}\eqnarray}%
  {\endeqnarray\stepcounter{equation}}
\makeatother

\newcommand{\exergue}[2]{\begin{flushright}{\it\scriptsize #1} {\bf\scriptsize #2}\end{flushright}}


\title{Moving Detectors in Cavities}

\author{Nathaniel Obadia}
\email{Nathaniel.Obadia@weizmann.ac.il}
\affiliation{Center for Astrophysics\\
The Weizmann Institute of Science\\
 Rehovot, Israel}
\pacs{03.70.+k, 04.62.+v, 04.70.Dy, 12.20.-m}

\begin{abstract}
We consider two-level detectors, coupled to a quantum scalar field, moving inside cavities.
We highlight some pathological resonant effects due to abrupt boundaries,
and decide to describe the cavity
by switching smoothly the interaction
by a time-dependent gate-like function.
Considering uniformly accelerated trajectories,
we show that some specific choices of non-adiabatic switching
have led to hazardous interpretations
about the enhancement of the Unruh effect in cavities.
More specifically, we show that
the emission/absorption ratio takes arbitrary high values according to the emitted quanta properties
and to the transients undergone at the entrance and the exit of the cavity,
{\it independently of the acceleration}.
An explicit example is provided where we show that inertial
and uniformly accelerated world-lines can even lead to the same
``pseudo-temperature''.
In passing, we also compute the deviation from the exact thermal response
for a finite size cavity.
\end{abstract}

\maketitle

\section{Introduction}

Quantum vacuum effects constitute
one of the most important classes of effects
in quantum field theory.
Among them,
because of the highlight it received from its analogy
with Hawking black hole radiation\cite{Hawking},
the Unruh effect\cite{Unruh} possesses a particular status.
It states that a linearly uniformly accelerating detector
with proper acceleration $a$ reacts, in the vacuum,
as if it was at rest in a thermal bath of quanta with temperature $\hbar a/2\pi ck_B$.
Many discussions arose and still arise
around the validity\cite{DeBievreMerkli,HuLin}
and the experimental feasibility\cite{Exp} of this issue.
In particular, whether or not some radiation is emitted is of much debate,
although the answer has already been clearly demonstrated twenty years ago\cite{Grove,RSG} and \cite{MaPa}.
Throughout the literature, the same crucial details are sometimes bypassed:
radiation emission and other peculiar effects due to non-adiabatic switching
are often melted to the (defined as above) Unruh effect,
which, because of the {\it uniform} character of the acceleration,
implies neither switching on nor off of the interaction.

However, this does not mean that such effects are irrelevant\cite{Pa,MaPa,Padma,HuLin}.
On the one hand, the study of non-adiabatic vacuum effects is interesting in itself.
On the other hand, the framework of the ``pure'' Unruh effect is far too idealistic,
so that one should consider some of its generalizations.
From an experimental angle,
switching on and off the interaction is unavoidable and its consequences have to be well understood,
in particular with respect to the radiation characteristics.

The aim of this article is
to clarify the physics of moving detectors in cavities
and eventually to correct one example
of the above mentioned questionable statements that have been proposed in the literature.
Namely, it has been claimed in \cite{Scully1,Scully2,Scully3} that,
by putting accelerated atoms in a single-mode semi-infinite cavity,
a significant enhancement of the emission/absorption ratio occurs:
from the thermal Unruh factor $e^{-2\pi E/a}$ it becomes $a/2\pi E$.
For accelerations that are small with respect to the detector proper frequency $E$,
but high enough to reach experimental scales,
the gain in temperature is claimed to be as high as a hundred orders of magnitude.
We correct this assertion
by pointing out that this new value of the emission/absorption ratio is no physically
linked to the Unruh effect,
to wit it is poorly correlated to the acceleration and it mainly depends on additional unquoted parameters.
More precisely, we will show that for a uniformly accelerated detector in a cavity,
1) one only has a pseudo-thermal population
since the ``temperature'' strongly depends on the quanta kinematical characteristics
(amplitude and direction),
and 2) the emission/absorption ratio significantly depends on the Doppler effect
undergone by the detector at the entrance and the exit of the cavity.
In other words, the origin of this phenomenon
lies in {\it transient effects} and not in {\it steady-state physics}.
Our conclusion is that, for a semi-infinite cavity
as treated in \cite{Scully1,Scully2,Scully3},
the corresponding value of the pseudo-temperature
associated to co-propagating quanta scales as $\gamma /\t_s^2k$,
$\gamma$ being the detector Lorentz factor at the entrance of the cavity,
$\t_s$ the switching proper-lapse,
and $k$ the quanta wavelength.

In order to deal with these issues,
we recall the usual two-level detector model
and some useful definitions in Section II.
Then, in Section III (as well as in Appendix A),
we look at non-adiabatic switching for inertial trajectories as a useful warm-up.
This enables us to shed some light
on the importance of properly defining the interaction
between the detector and the quantum field.
Section IV presents the results
for uniformly accelerated motions
and introduces a useful and well-adapted switching function.
The latter allows us to deal with semi-infinite cavity specificities in Section V.
We conclude in Section VI.
Finally, in Appendices B and C,
we review the intricate mathematical properties of the transition amplitudes.

\section{The settings}

We consider a two-level detector ($\ket{-},\ket{+}$), defined by its energy gap $E$,
that moves along a time-like trajectory in $3+1$ Minkowski space-time.
This detector is coupled to a massless scalar field $\Phi$:
\ba\label{Phi}
\Phi(x)= \inte \! \frac{\mbox{d} \vec{k}}{\sqrt{2{(2\pi)}^3k}} \;
\left(
a_{\vec{k}} \, e^{-i k_\mu  x^\mu}
\ + \
a_{\vec{k}}^\dagger \, e^{+i k_\mu  x^\mu} 
\right)
\ea
where $a_{\vec{k}}^\dagger$ and $a_{\vec{k}}$ are the creation and annihilation operators
obeying the usual commutation rules
$\left[ a_{\vec{k}},a_{\vec{k'}}^\dagger\right] = \delta(\vec{k}-\vec{k'})$.
The interaction is localized along the detector trajectory $x^\mu(\t)$,
where $\t$ is the proper-time.
For matter of simplicity, the coupling is chosen to be linear.
In order to control its switching on and off,
the interaction Hamiltonian contains an explicitly time dependent function $f(\t)$:
\ba\label{Hint}
H(\t) = g \: f(\t) \: \left(
e^{iE\t} \, \ket{+}\bra{-} \: + \: e^{-iE\t} \, \ket{-}\bra{+}
\right) \;  \Phi(x^\mu(\t)) 
\ea
where $g$ is an a-dimensional coupling constant that is assumed to be small.
This constant is chosen such that $f(\t)$ takes values between $0$ and $1$.
For instance, mimicking a perfect cavity necessitates
a sharp finite-time interaction, beginning at $\t_i$ and ending at $\t_f$:
$f_{G}(\t)=\Theta(\t-\t_i) \Theta(\t_f-\t)$, where $\Theta$ is the Heaviside function.

If one assumes that the total system was in its vacuum state $\vac$,
and that the detector was in its lowest state $\ket{-}$
in the infinite past,
then the state of the system at any proper-time $\t$ is given by:
\ba\label{Psi}
\ket{\Psi_-(\t)}
=\hat T \: e^{\disp -i\,\int^\t_{-\infty} \! \mbox{d}\t' \; H(\t)} \; \ket{0,-} \ ,
\ea
when $\hat T$ means time-ordering.
In general, since one controls both the beginning and the end of the interaction through $f(\t)$,
one can focus on the sole object $\ket{\Psi_-(\t=+\infty)}$, while varying the properties of $f$.
For simplicity, we will denote this ket $\ket{\Psi_-^\infty}$ in the remainder of the paper.
In this way, our approach is similar to that of a scattering process.

The physics of such moving detectors depend on two degrees of freedom:
the choice of the detector's trajectory $x^\mu(\t)$,
and the time-dependent character of the interaction $f(\t)$.
In this article, we will only focus on inertial and linearly uniformly accelerated trajectories.
Moreover, since we want to mimic interaction in cavities, we will not consider any fuzzy $f(\t)$
but rather gate-like switching functions.
The basic tools for the description of the interaction are the transition amplitudes:
the amplitude to create a particle of momentum $\vec{k}$ and simultaneously to raise the detector level is
\ba
A_{+,\vec{k}}(E) &\equiv& \left(\bvac a_{\vec{k}} \otimes \bra{+}\right)\ket{\Psi_-^\infty} \label{Amplitude}\\
&=& \frac{-i g}{\sqrt{2{(2\pi)}^3k}}  \Ie{\t} f(\t) \: e^{i[E\t + {k}_\mu x^\mu(\t)]}\label{Amplitude2}
\ea
and its counterpart $A_{-,\vec{k}}\equiv\left(\bvac a_{\vec{k}} \otimes \bra{-}\right)\ket{\Psi_+^\infty}$,
which corresponds to a lowering of the detector level when the initial state is $\ket{+}$,
is obtained by replacing $E$ by $-E$ in \reff{Amplitude2}.
(As we focus on the first order in the perturbation expansion, we are not interested in amplitudes such as
$\left(\bvac a_{\vec{k}} a_{\vec{k'}}\otimes \bra{-}\right)\ket{\Psi_-^\infty}$ for instance.)
When using the projectors over the subspaces containing one $\vec{k}-$particle, 
$\mathbb{P}_{\pm,\vec{k}}\equiv \left(a_{\vec{k}}^\dagger\vac\otimes\ket{\pm}\right) \Big(\bra{\pm}\otimes\bvac a_{\vec{k}}\Big)$,
the transition probabilities to create a quanta $\vec{k}$ and to raise (lower) the detector level
naturally involve the previous amplitudes:
\ba\label{defPveck}
P_{\pm,\vec{k}}(E) &\equiv& \expect{\Psi_\mp^\infty}{\mathbb{P}_{\pm,\vec{k}}}{\Psi_\mp^\infty}
= \left|A_{\pm,\vec{k}}\right|^2 \ .
\ea
Similarly, the probability to create a particle of energy $|\vec{k}|$ (whatever its momentum direction)
and to raise (lower) the detector level
is
\ba\label{defPk}
P_{\pm,k=|\vec{k}|}(E) &\equiv& \Iinfsup{0}{2\pi}{\varphi} \Iinfsup{0}{\pi}{\theta} \sin\theta \: P_{\pm,\vec{k}} \ .
\ea
Finally, the probability to raise (lower) the detector level independently of the particle created is
\ba\label{defP}
P_{\pm}(E) &\equiv&  \Io{k} k^2 \:  P_{\pm,k} \ .
\ea
The previous objects allow to construct the $g-$independent ratios defined by
\ba
r_{\vec{k}}(E)\equiv\frac{P_{+,\vec{k}}}{P_{-,\vec{k}}} \; ,
r_{k}(E)\equiv\frac{P_{+,k}}{P_{-,k}} \; ,
r(E) \equiv\frac{P_{+}}{P_{-}} \ .
\ea
When steady-state is reached,
one assumes that the principle of detailed balance holds\cite{LesHouchesdeSitter},
and the latter ratios reflect the repartition between the ground state and the excited one's populations: $n_+/n_-=P_+/P_-$.
As a common feature to all trajectories, when the gap $E$ tends to zero,
then the emission and absorption processes become equivalent,
their corresponding amplitudes merge, and all the previous ratios tend to one
if they are mathematically well-defined.
In addition, the ratios $E-$dependence may exhibit interesting properties,
such as exponential decay for uniformly accelerated motions.
This will be the subject of Sections IV and V.

Before entering into these details, let us stress the specific situation we wish to consider.
Here we are interested in moving detectors in cavities.
We will focus on linear motions in the increasing $z$ direction along the $z$ axis.
We will assume that the cavity respects a cylindrical symmetry around this axis
such that the integration over $\varphi$ trivially reduces to a $2\pi$ multiplying factor.
Finally, $\theta$ is the angle between the particle's momentum $\vec{k}$ and $\vec{z}$.

Moreover, let us be clear about the role of the switching function:
it is made to mimic a cavity that the detector enters at some proper time $\t_i$ and leaves at $\t_f$.
To this end, $f(\t)$ takes values close to $1$ in the interval $[\t_i, \t_f]$
and decreases rapidly to $0$ outwards.

\section{Inertial motion}

In order to get a better understanding of the uniformly accelerated motion physics,
we first inspect the outcomes of an inertial trajectory.
It is usually admitted that inertial motion does not lead to any particle creation process.
However this statement is only true in a steady state regime.
Indeed, transient effects generated by the switching on and off of the interaction
are sufficient to put on mass shell a ``pair'' of particles,
namely a real $\Phi$-particle accompanied by a (des-)exciton of the detector:
\ba
A_{\pm,\vec{k}}^{I}(E) &=& \frac{-i g}{\sqrt{4\pi k}} \; f_{\pm E+k \xi(\theta)} \ ,
\ea
where $f_\om=\Ie{\t}f(\t)e^{i\om\t}/2\pi$ is the Fourier component of $f(\t)$.
The Doppler factor is
$\xi(\theta)\equiv \gamma (1-\vec{v}.\vec{k}/k)=\gamma(1-v\cos\theta)\in[e^{-\nu},e^{\nu}]$,
where $\vec{v}=\vec{z}\tanh\nu $ is the
detector's constant speed, $\nu>0$ is the rapidity and $\gamma=1/\sqrt{1-v^2}=\cosh\nu$ the Lorentz
factor.
As specific examples,
a co[counter]-propagating particle $(\theta=0[\pi])$ has $\xi=e^{-\nu}[e^{\nu}]$,
a transverse particle $(\theta=\pm\pi/2)$ has $\xi=\cosh\nu$.
Note that only co-propagating quanta can undergo arbitrary small Doppler effects;
this particularity will play an important role in the following.

Before probing transient effects, we recollect that for a uniform coupling, $f(\t)\equiv 1$,
the emission amplitude identically vanishes
$A_{+,\vec{k}}^{I} \propto \delta(E+k \xi(\theta)) \equiv 0$,
whereas the absorption amplitude is non zero only at resonance $A_{-,\vec{k}}^{I} \propto \delta(E-k \xi(\theta))$.

More generally, for any switching function, the emission/absorption ratios are straightforwardly given by
\ba
r^{I}_{\vec{k}}(E)\label{rinertveck}
&=& \left| \frac{f_{E+k \xi(\theta)}}{f_{-E+k \xi(\theta)}} \right|^2 \ ,\\
r^{I}_{k}(E)\label{rinertk}
&=&  \frac{\Iinfsup{e^{-\nu}}{e^{\nu}}{x}\left|f_{E+kx}\right|^2}{\Iinfsup{e^{-\nu}}{e^{\nu}}{x}\left|f_{-E+kx}\right|^2} \ ,\\
r^{I}(E)\label{rinert}
&=&  \frac{\Io{k}k\Iinfsup{e^{-\nu}}{e^{\nu}}{x}\left|f_{E+kx}\right|^2}{\Io{k}k\Iinfsup{e^{-\nu}}{e^{\nu}}{x}\left|f_{-E+kx}\right|^2}\nn
&=&\frac{\Io{k}k\left|f_{E+k}\right|^2}{\Io{k}k\left|f_{-E+k}\right|^2}\ ,\label{rinert2}
\ea
where the velocity dependence explicitly vanishes in \reff{rinert2}
since the integration over all the possible modes ensures the ratio $r^{I}$ to be a Lorentz invariant.

Describing a cavity can obviously be done by choosing a gate function
(for an extensive study, see {\it e.g.} \cite{Padma}).
Thereby, it is foreseeable that abrupt boundaries generate resonant effects
that give extreme values to the amplitudes, and consequently to the ratios.
We address this issue for completeness but, since this discussion somehow lies out of the scope of this paper,
we develop it in Appendix A.
Our main conclusion is that perfect boundaries should be discarded
and switching functions have to be carefully chosen in order
to avoid resonant effects and non-physical behavior of the amplitudes.

As our aim is to understand the physics of uniformly accelerated detectors,
note that a mere gaussian switching function provides exactly a ``thermal'' answer in the inertial case:
\ba
f_{g}(\t) &=& e^{-\disp (\t-\t_0)^2/2\Lambda^2} \nn
&& \Rightarrow \;
{}^{g}r^{I}_{\vec{k}} = e^{-\disp 4 \Lambda^2 k \xi(\theta) E} \ ,\label{rinertveckgauss}
\ea
which corresponds to a ``temperature'' $T_{g}=[4\Lambda^2 k \xi(\theta)]^{-1}$.
If the gaussian function mimics accurately the intensity of the interaction, 
then $\Lambda$ is roughly equal to its proper duration $L/\gamma v$ around $\t_0$.
Therefore, this pseudo-temperature is expressed in terms of the laboratory's frame quantities and scales as $\sinh^2\nu/kL^2\xi(\theta)$.
For given cavity length $L$ and quanta wave-vector  $k$,
counter-propagating or transverse modes coupled to sufficiently relativistic detectors provide $T_{g}\propto \gamma/L^2k$,
whereas co-propagating modes give a hotter component to the bath $T_{g}\propto \gamma^3/L^2k$.
The latter is the consequence of the fact that only co-propagating quanta can undergo
arbitrary small Doppler effects.

When integrating over $\theta$, the exact thermal feature disappears since the temperature in \reff{rinertveckgauss} is $\theta-$dependent.
However, one recovers a pseudo-thermal behavior for high energies since
\ba\label{rinertkgauss}
{}^{g}r^{I}_{k} = \frac{\erf[\Lambda(E+ke^{\nu})]-\erf[\Lambda(E+ke^{-\nu})]}
{\erf[\Lambda(E-ke^{-\nu})]-\erf[\Lambda(E-ke^{\nu})]}
\ea
behaves as $e^{-4\Lambda^2 k e^{-\nu} E}$ when $\nu$ and $E$ are sufficiently large, namely $E\gg ke^\nu\gg ke^{-\nu}$.
($\erf(x)=2\Iinfsup{0}{x}{t}e^{-t^2}/\sqrt{\pi}$ is the error function.)
The integration keeps track of the hottest (co-propagating) modes and, therefore, the pseudo-temperature also scales as $T_{g}\propto \gamma^3/L^2k$.
Notwithstanding, any pseudo-thermal behavior vanishes in the $k-$integrated ratio since one obtains
\ba\label{rinertgauss}
{}^{g}r^{I} = \frac{1-\sqrt{\pi}\Lambda E(1-\erf[\Lambda E])e^{\Lambda^2E^2}}{1+\sqrt{\pi}\Lambda E(1+\erf[\Lambda E])e^{\Lambda^2E^2}}
\ea
which scales as $e^{-\Lambda^2E^2}/(\Lambda E)^3$ for large energies.

We addressed this academic issue to reach the following statement:
According to the last three equations, one would {\it not} conclude that inertial detectors in a gaussian cavity are thermally populated
or that they undergo the Unruh effect.
One would rather interpret \reff{rinertveckgauss} as a kind of artefact, noticing
that \reff{rinertkgauss} is just asymptotically thermal and \reff{rinertgauss} not at all.
We will somehow apply a similar line of thought to uniformly accelerated detectors in order to reach the same type of conclusion.

Before devoting ourselves to uniformly accelerated motion,
it is interesting to point out that in the previous study the choice of $\t_i$ is irrelevant as long as $\t_f-\t_i$ is kept fixed.
This is obviously due to the fact that the Doppler shift is constant along the trajectory for inertial trajectories;
this will no longer occur in the next sections when looking at uniformly accelerated motions.

\section{Uniformly accelerated linear motion}

Uniform acceleration is  undergone by world-lines that obey
$a^\mu a_\mu\equiv\frac{\d^2\disp x^\mu}{\d\disp\t^2}\frac{\d^2\disp x_\mu}{\d\disp \t^2}=-a^2$.
When requiring that the proper-time along the trajectory is oriented as the Minkowski time,
one obtains, for instance, the hyperbolic trajectory $x^\mu(\t)=(\sinh(a\t)/a,0,0,\cosh(a\t)/a)$.

The emission and the absorption amplitudes are no longer trivially proportional to some Fourier components of $f(\t)$ but are given by:
\bwt
\ba\label{Aunifacccpmgeneral}
A_{\pm,\vec{k}}^{UA}&=& \frac{- i g}{\sqrt{2{(2\pi)}^3k}} \;
\Ie{\t} f(\t) \: e^{i (\pm E \t + k(1-\cos\theta) e^{a\t}/2a - k(1+\cos\theta) e^{-a\t}/2a )} \ .
\ea
\ewt

\subsection{The Unruh effect}

First, we recover the celebrated Unruh result for a uniform interaction, $f(\t)\equiv 1$.
In Appendix B1, we reveal how \reff{Aunifacccpmgeneral} can be re-written in terms of $K$ modified Bessel functions\cite{Abra},
for any off-axis quanta ($\cos\theta\neq\pm 1$), so as to obtain
\bwt
\ba\label{Aunifaccpmoffaxis}
{}^{U}A_{\pm,\vec{k}}^{UA}&=& \frac{- 2 i g}{\sqrt{2{(2\pi)}^3k}} \;
\frac{e^{\mp\pi E/2a}}{a} {\left(\frac{1+\cos\theta}{1-\cos\theta}\right)}^{\pm iE/2a} \: K_{\pm iE/a}\left[\frac{k\sin\theta}{a}\right] \ .
\ea
\ewt
Since $K_\nu(z)=K_{-\nu}(z)$ for any $(\nu,z)$, one finds that the emission/absorption ratio is given by
\ba\label{Unruhratio}
{}^{U}r^{UA}_{\vec{k}}
=e^{\disp -2\pi E/a} \ .
\ea
When $\cos\theta=\pm 1$, the emission and absorption amplitudes differ from \reff{Aunifaccpmoffaxis};
they are expressed in terms of Gamma functions.
However, the final outcome is identical, see Appendix B2, and
on-axis quanta provide the Unruh result \reff{Unruhratio}, similarly to off-axis ones.

The main particularity of \reff{Unruhratio}
is that it is independent of the nature of the particles created,
both of the value of their wave-vector $k$, and of their direction $\theta$,
although the detector possesses two dimensionful scales $a$ and $E$ and a preferred direction $\vec{z}$.
Indeed, \reff{Aunifaccpmoffaxis} provides $\left|{}^{U}A_{\pm,\vec{k}}^{UA}\right|^2=e^{\mp \pi E/a} \times F(\vec{k},|E|)$.
Therefore, the other two ratios $({}^{U}r^{UA}_k,{}^{U}r^{UA})$
also reduce to the thermal factor $e^{-2\pi E/a}$.
From the constancy of these ratios, and when assuming the detailed balance principle,
one finds that the energy levels of the system are thermally populated with temperature $T_U=a/2\pi$\cite{Unruh}.
Then, one deduces that a uniformly accelerated two-level detector feels the vacuum
as if it was a thermal bath at temperature $T_U$.
Note also that if we had taken photons instead of scalars,
the index of the Bessel functions would have been replaced by $1\pm i E/a$
and \reff{Unruhratio} would still hold since
$[K_\nu(z\in\mathbb{R})]^*=K_{\nu^*}(z\in\mathbb{R})$.
\cite{footnotereftoscully3}

\subsection{Mimicking a cavity}

In order to mimic a cavity, one could impose abrupt boundaries as we did in Appendix A.
However, we know that such a choice leads to pathological values of the ratios, liable to resonance effects.
Therefore, we propose the following switching function\cite{OPa2}:
\ba\label{superf}
f(\t)=e^{-\eta_1 e^{-a\t}-\eta_2 e^{a\t}} \ .
\ea
\reff{superf} exhibits a plateau of value $\sim 1$
which begins around $\t_i\equiv\ln\eta_1/a$ and finishes around $\t_f\equiv -\ln\eta_2/a$.
The switching duration is a few $1/a$, independently of the $\eta$'s in the $\eta_1\eta_2\ll 1$
regime\cite{footnotef}.
Due to the hyperbolic character of the uniformly accelerated trajectory,
the movement points towards the increasing $z$ for $\t>0$ only, since $v=\tanh a\t$.
Therefore, mimicking injected detectors in a cavity necessitates $\t_i>0 \Leftrightarrow \eta_1>1$.
The exiting time is related to the length $L$ of the cavity by $\cosh a\t_f=aL+ \cosh a\t_i$.
Thus the condition $\eta_1\eta_2\ll 1$ implies $L\gg 1/a$.
The usefulness of \reff{superf} lies in the fact that this function is adjusted to the trajectory,
since its only dimensional quantity is precisely the acceleration $a$.
As a useful consequence, it preserves the ability to express the amplitudes in terms of the same Bessel functions:
\bwt
\ba
{}^{cav}A_{\pm,\vec{k}}^{UA}&=& \frac{- 2i g}{\sqrt{2{(2\pi)}^3k}} \;
\frac{e^{\mp \frac{E}{2a}\arctan(k_+/\eta_1) \mp \frac{E}{2a}\arctan(k_-/\eta_2)}}{a} \:
{\left(\frac{{\eta_1}^2 + {k_+}^2}{{\eta_2}^2 + {k_-}^2}\right)}^{\pm iE/4a} \nn
&& \; \; \times \; K_{\pm iE/a}\left[2{({\eta_1}^2 + {k_+}^2)}^{1/4}{({\eta_2}^2 + {k_-}^2)}^{1/4} \:
e^{\frac{i}{2}\arctan(k_+/\eta_1) - \frac{i}{2}\arctan(k_-/\eta_2)} \right] \ ,\label{Aunifaccsmoothoffaxis}
\ea
\ewt
where $k_\pm\equiv k(1 \pm \cos\theta)/2a$.
Detailed computations are shown in Appendix C1, where we also stress that
\reff{superf} is not an arbitrary choice since it encodes the transition amplitudes analytical properties.

Comparing \reff{Aunifaccsmoothoffaxis} with Eqs.(\ref{Aunifaccpmoffaxis}-\ref{Unruhratio}), one finds that,
for any off-axis process ($k_+k_-\neq 0$), the emission/absorption ratio is given by
\bwt
\ba\label{ratiooffaxissmooth}
{}^{cav}r^{UA}_{\vec{k}}
&=& e^{\disp-\frac{2E}{a}\disp\left(\arctan\left[\frac{k(1 + \cos\theta)e^{-a\t_i}}{2a}\right] \: + \:
\arctan\left[\frac{k(1 - \cos\theta)e^{a\t_f}}{2a}\right]\right)} \ .
\ea
\ewt
\reff{ratiooffaxissmooth} is one of our main results and calls for some comments.
First, if one wants to retrieve exactly the Unruh result, the interaction has to be uniform,
\ie one has to take the double limit $\t_i\to -\infty,\t_f\to +\infty$.
Both $\arctan$ tend to $\pi/2$ and the amplitudes \reff{Aunifaccsmoothoffaxis} (resp. the ratio \reff{ratiooffaxissmooth})
give back the Unruh amplitudes \reff{Aunifaccpmoffaxis} (resp. the Unruh ratio \reff{Unruhratio}).
However, \reff{ratiooffaxissmooth} also allows to estimate the convergence to the Unruh ratio when
the lapse $\Delta\t=\t_f-\t_i$ becomes large.
For instance, for transverse quanta, the relative correction with respect to \reff{Unruhratio}
behaves as
\ba\label{relativecorrection}
\frac{{}^{cav}r^{UA}_{\vec{k}}-{}^{U}r^{UA}_{\vec{k}}}{{}^{U}r^{UA}_{\vec{k}}}
\simeq \frac{8E}{k} \: \cosh\left(\frac{a(\t_f+\t_i)}{2}\right) \: e^{-\disp\frac{a\Delta\t}{2}} \  .
\ea
We believe that the exponential decay with respect to the proper lapse
is generically reobtained after resummation over all the possible modes.
Indeed, $\Delta\t$ is the only Lorentz invariant of the problem and
boosts along the uniformly accelerated motion correspond to proper-time translations.
Since we chose the switching function (\ref{superf}) such that it preserves the mathematical
properties of uniformly accelerated physics, this property should still hold.
Moreover, a similar exponential decay to the perfect Unruh picture has already been
found in a closely related previous work (see Eqs.(47) and (55) in \cite{OPa2}).

Second, the cautious reader may wonder why the limit $\t_i\to\t_f$ does not provide a trivial result.
The answer lies in the fact that $\t_{i,f}=\pm\ln\eta_{1,2}/a$ reliably
represent the switching on and off times only in the limit $\t_f-\t_i\gg 1/a$, see \cite{footnotef}.

Third, \reff{ratiooffaxissmooth} means that the pseudo-temperature obeys
\ba\label{betasmooth1}
1/T^{cav} &=& 1/T^{cav}_i + 1/T^{cav}_f < \frac{2\pi}{a}\ ,\\
T^{cav}_{i,f} &=& \left( \frac{2}{a}\:\arctan\left[k(1\pm\cos\theta)e^{\mp a\t_{i,f}}/2a\right] \right)^{-1}\ .\label{betasmooth2}
\ea
Hence, the pseudo-temperature is always larger than the Unruh temperature.
This enhancement is due to the fact that mimicking a cavity induces transients that ``over-heat'' the detector,
contrary to the Unruh picture whose temperature is generated by steady-state coherent processes.
This interpretation goes with the results found by looking at the non-vanishing Rindler flux during such processes\cite{MaPa}.
We refer again to a ``pseudo-temperature'' since, contrary to \reff{Unruhratio},
$T^{cav}$ depends on $\vec{k}$, which prevents from recovering thermal expressions for
$r^{UA}_{k}$ and $r^{UA}$, as we saw in the inertial case, Eqs.(\ref{rinertkgauss}-\ref{rinertgauss}).

\begin{figure}[h!]
\hbox to\linewidth{\hss
\resizebox{7.6cm}{5.6cm}{\includegraphics{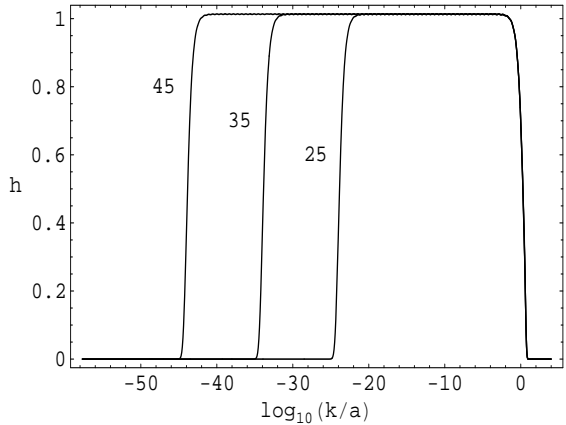}}
\hss}
\caption{The energy density $h$ (arbitrary units) as a function of $\log_{10}(k/a)$
for different values of $\log_{10}(a L)$ (as marked).
The gap is $E/a=5$ and the switching on time corresponds to $\eta_1=2$.\label{Fig5}}
\end{figure}

\begin{figure}[h!]
\hbox to\linewidth{\hss
\resizebox{7.6cm}{5.6cm}{\includegraphics{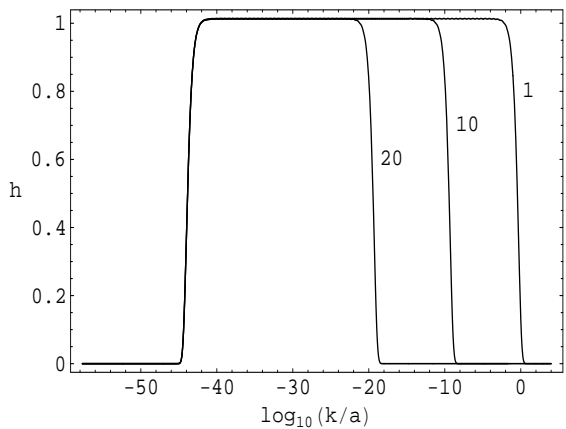}}
\hss}
\caption{The energy density $h$ (arbitrary units) as a function of $\log_{10}(k/a)$
for different values of $\log_{10}\eta_1$ (as marked).
The gap is $E/a=5$ and the cavity length is $L=10^{45}/a$.\label{Fig6}}
\end{figure}

Fourth, a noticeable feature of Eqs.(\ref{betasmooth1}-\ref{betasmooth2}) is that the pseudo-temperature is unbounded from above
in the $k\to 0$ limit since it scales as $T^{cav}\propto [a^2/((1+\cos\theta)/\eta_1+(1-\cos\theta)/\eta_2)]/ k$.
However, one can legitimately question the meaning of considering extremely large wavelengths in finite size cavities.
Indeed, one should expect only a subset of modes to survive in a cavity.
In order to describe how the spectrum is modified, let us use the amplitudes WKB expressions
(see \cite{primer,MaPa,OPa2,OPa3} for various applications of this method).
By inspecting \reff{Aunifacccpmgeneral} using \reff{superf} for transverse quanta,
the integrand presents a saddle time which obeys $\re(a\t^*(k))\simeq \ln (E/k)$
in the $\eta_1>1, \eta_1\eta_2\ll 1$ regime.
Therefore only the modes such that $0<\t_i<\re(\t^*(k))<\t_f$ are allowed within the cavity.
This reduces to
\ba
\inv{2 a L } \lesssim\frac{k}{E} \lesssim \inv{\eta_1}
\ea
which we illustrate in Figs.(\ref{Fig5}) and (\ref{Fig6})
by computing the energy density for the de-excitation process
$h=k |{}^{cav}A_{-,\vec{k}}^{UA}|^2$.
By extending the size of the cavity, one allows the long wavelengths to be created.
However, since $\t_i>0$, only $k<E$ modes are allowed in any case.
All modes can participate if one allows the detector to "come and go",
\ie if $\t_i \to - \infty$.

\begin{figure}[h!]
\hbox to\linewidth{\hss
\resizebox{7.6cm}{5.6cm}{\includegraphics{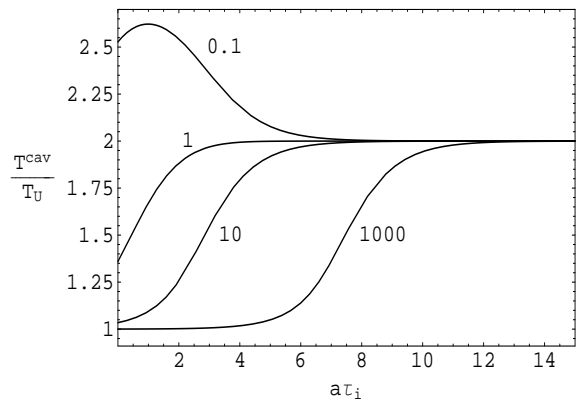}}
\hss}
\caption{The ratio $\frac{T^{cav}}{a/2\pi}$ as a function of $a\t_i$
for different values of $k/2a$ (as marked).
We consider mode-blocked transverse quanta $(\theta=\pi/2)$ and a fixed cavity length $L=10/a$.
When $\t_i\to\infty$, $\beta_i\to 0$ and $\t_f\to\infty$ causes $\beta_f\to \pi/a$ in \reff{betasmooth1}
which explains why $\frac{T^{cav}}{a/2\pi}\to 2$.
In all cases, the pseudo-temperature is higher than the Unruh temperature.
It can be arbitrary high for small mode energies $k$ and early switching on times $\t_i$.\label{Fig1}}
\end{figure}

\begin{figure}[h!]
\hbox to\linewidth{\hss
\resizebox{7.6cm}{5.6cm}{\includegraphics{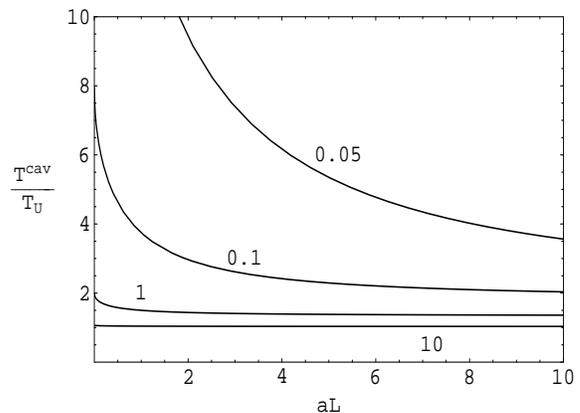}}
\hss}
\caption{The ratio $\frac{T^{cav}}{a/2\pi}$ as a function of $a L$
for different values of $k/2a$ (as marked) when $\theta=\pi/2$ and $\t_i=0$.
When $L\to\infty$, $\beta_f\to 0$ and $\beta\to\beta_i$ in \reff{betasmooth1}.
In all cases, the pseudo-temperature is higher than the Unruh temperature.
It can be arbitrary high for small enough mode energies $k$.
The always decreasing character of $T^{cav}(L)$ can be understood directly from the fact that
$\beta_f$ is an increasing function of $\t_f$, therefore of $L$.
For very large cavities, the pseudo-temperature tends to $1/\beta_i=a/2\arctan(k/2a)$.
\label{Fig2}}
\end{figure}

For a mode-blocked cavity, when considering off-axis quanta,
the value of this pseudo-temperature is dictated by transient effects,
\ie by the switching times ${\t_{i,f}}$.
We illustrate the dependence of $T^{cav}$ with respect to $\t_i$ and $L$ in Figs.(\ref{Fig1}) 
and (\ref{Fig2}).

From Eqs.(\ref{betasmooth1}-\ref{betasmooth2}) one learns that the ``temperature'' is
{\it always} larger than the Unruh value because of the switching effects.
However, a greater enhancement of the ratios, independent of $k$,
has been claimed in the literature; the corresponding situation is the subject of the next section.


\section{Semi-infinite cavity}

In this section, we focus on the physics obtained when accelerating a detector
from a given initial moment $\t_i$ and for a long time $\t_f-\t_i \ggg 1/a,1/E,1/k$.
This situation is formally obtained by putting $\eta_2$ to zero,
while letting $\eta_1$ take any value larger than unity.

It has been claimed by Scully {\it et al.}\cite{Scully1,Scully2,Scully3} that,
by putting accelerated atoms in a single-mode cavity,
a significant enhancement of the "radiation" is produced
such that the emission/absorption ratio becomes $a/2\pi E \ll 1$
instead of the thermal Unruh factor $e^{-2\pi E/a}\lll 1$.
We would like to comment on this, with the help of our previous expressions.
Hu and Roura\cite{HuRoura} contested such a result
by arguing that this great enhancement appears in a regime in which the acceleration no longer plays a crucial role.
However, they did not give any explicit proof to support this claim.
The answer provided by the authors of \cite{Scully1} to this objection is that the ratio being proportional to the acceleration,
the latter quantity can hardly be called unrelated \cite{Scully2}.
We want to remark that, in the treatment of \cite{Scully1,Scully2,Scully3}, apart from $k$,
which is chosen to be much larger than the acceleration $a$ and the energy $E$,
the latter two values are the only dimensional quantities.
Therefore, any final a-dimensional outcome must depend on $a/E$.

According to us, the very expression $r_{\vec{k}}=a/2\pi E$ is in no way a brand new cavity-induced Unruh effect.
There are several (related) reasons for this.

1) Such a value of the ratio means a strongly energy-dependent expression $T\simeq E/\ln(2\pi E/a)$,
which, hence, can hardly be called ``temperature''.
As a comparison, such is not the case in the so-called ``circular Unruh effect''\cite{BellLeinaas,UnruhCircular,OMi},
which provides a poorly $E-$dependent temperature $T=E/\ln(1+4\sqrt{3}e^{2\sqrt{3}E/a}E/a) \in [a/4\sqrt{3},a/2\sqrt{3}]$.

2) It is difficult to understand how the presence of a cavity can transform
a situation for which {\it all} the detectors undergo the {\it same} thermal process (the original Unruh effect)
into a picture that describes a {\it detector-dependent} and  {\it cavity-independent}
response, since $T$ depends on $E$ and not on the cavity characteristics.\cite{footnotevisser}

3) Finally, $a/2\pi E$ is {\it not at all} a maximal value
of the emission/absorption ratio, as we reveal by examining the precise consequences of the Doppler effect
in the same situation as studied in \cite{Scully1}, that is when $\theta=0$.

The emission/absorption ratio for co-propagating quanta in a semi-cavity is
\ba\label{ratiosemiinfonaxis}
{}^{s.cav}r^{UA}_{\vec{k}} = e^{\disp -\frac{4 E}{a}\arctan(k e^{-a\t_i}/a)}  \ .
\ea
(This result is not trivially obtained from \reff{ratiooffaxissmooth} since the latter contains the expression $k_-/\eta_2$
where both $k_-$ and $\eta_2$ are formally vanishing for co-propagating quanta in semi-infinite cavities;
details are provided in the Appendix C2.)
Note that, in the limit of uniform coupling, $\t_i\to -\infty \Leftrightarrow \eta_1\to 0$,
one recovers the Unruh result since $r^{UA}_{\vec{k}}\to e^{-2\pi E/a}$.

\reff{ratiosemiinfonaxis} does not suffer from problems 1) and 2) since
the pseudo-temperature $T^{s.cav}=a/4\arctan(k e^{-a\t_i}/a)$ is energy-independent and cavity-dependent.
More precisely, the ratio does not depend on the {\it lapse} during which the interaction lasts (since it is infinite)
but on the {\it moment} when it begins, through the value of the parameter $\eta_1$:
this is a straightforward signature of the transients undergone at the entrance of the cavity.

\begin{figure}[h!]
\hbox to\linewidth{\hss
\resizebox{7.6cm}{7cm}{\includegraphics{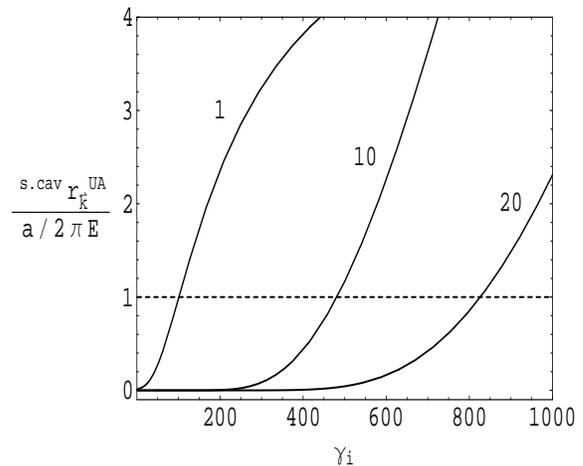}}
\hss}
\caption{The ratio ${}^{s.cav}r^{UA}_{\vec{k}}/(a/2\pi E)$ for co-propagating quanta
as a function of the Lorentz factor at the entrance of the cavity $\ga_i=\cosh(a\t_i)$
when $k/a=100$ and for different values of $E/a$ (as marked).
All curves go beyond unity at one point, depending on the values of $k/a$ and $E/a$.
\label{Fig4}}
\end{figure}

\reff{ratiosemiinfonaxis} also addresses point 3).
When ``waiting'' enough, \ie when assigning large values to $\eta_1=e^{a\t_i}$ with respect to unity,
the emission/absoption ratio can be put as high as one wishes while preserving $r<1$.
Since $v=\tanh a\t$, requiring large values of $\eta_1=\sqrt{(1+v_i)/(1-v_i)}$
is equivalent to injecting the detector with a higher velocity in the cavity, \ie large $\ga_i$'s.
In this case, the pseudo-temperature $T^{s.cav}\propto \gamma_i a^2/k$ is proportional to the square of the acceleration,
and, as a sign of its non-adiabatic origin, to the Lorentz factor at the entrance of the cavity in the high $\gamma_i$ regime.
Therefore, $a/2\pi E$ is just an intermediate value obtained for some initial time $\t_i(r=a/2\pi E) \simeq \ln\left[4kE/a^2\ln(2\pi E/a)\right]/a$,
and constitutes in no way an upper limit for $r^{UA}_{\vec{k}}$
since the latter takes, as a function of $\t_i$, all values between $e^{-4E\arctan(k/a)/a}$ and $1$,
see Fig.(\ref{Fig4}).

The precise value of $\t_i(r=a/2\pi E)$ depends on our model,
that is, on our choice of the switching function \reff{superf}.
Had we chosen another $f(\t)$ that allows analytical expressions, we would have found another value of
it\cite{footnoteWKB}.
However, the effect itself, that is the Doppler boosted transient enhancement, is a common feature for all (gate-like) $f(\t)$.
In order to grasp the reason why $\t_i(r=a/2\pi E)=0$ in \cite{Scully1,Scully2,Scully3},
note that their model is made out of abrupt boundaries
(\ie perfect gate switching functions that we know to be pathological after Section II and Appendix A),
and approximated incomplete Gamma functions.
Concerning the absence of $\t_i-$dependence in the ratios found in \cite{Scully1,Scully2,Scully3},
one should notice that the authors chose to get rid of the two boundaries dependence in two different ways:
by putting $\t_f=+\infty$ as we did, and by assigning $\t_i=0$ {\it a priori}, which obviously masked its effect.
Therefore, as we already pointed out, their final ratio could not depend clearly on any undergone Doppler effect at the entrance of the cavity.

What is important is to notice that the main quantity that drives the value of the emission/absorption
ratio for co-propagating quanta is the (exponentially small) Doppler effect exerted on emitted quanta
at the onset of the interaction.
The later the switching on,
the smaller the Doppler factor,
the more red-shifted the emitted quanta,
the hotter the detector,
and the larger the emission/absorption ratio.

Finally, note that the preceding results are consistent with what we learned in Section III.
Indeed, by looking at co-propagating quanta for a ultra-relativistic constant speed detector in a gaussian cavity,
we found that \reff{rinertkgauss} provides the pseudo-temperature $T^{g}\propto \gamma^3/L^2k$.
Knowing that the undergone effect is due to the {\it switching} and not to the {\it cavity length},
and that both switching duration $\t_s$ and interaction size $L/\gamma v$ scale as $\Lambda$ when choosing a gaussian profile,
one can rewrite the pseudo-temperature as $T^{g}\propto \gamma/\t_s^2k$.
Now, noticing that the switching time $\t_s$ is precisely $1/a$ in the accelerated case\cite{footnotef},
$T^{s.cav}\propto \gamma_ia^2/k$ can also be rewritten as $T^{s.cav}\propto \gamma_i/\t_s^2k$.
Therefore, the generic expression relating the pseudo-temperature $T_k$ of a detector
sensible to co-propagating particles of momentum $k$,
to the Lorentz factor at the entrance of the cavity $\ga_s$,
and to the proper switching lapse $\t_s$,
can be obtained {\it similarly for the inertial and the uniformly accelerated case}
in the high $\ga_s$ regime:
\ba T_k \propto \gamma_s /\t_s^2k \ .\label{TheRelation}\ea
\reff{TheRelation} prompts to follow Hu and Roura
when they state ``when the emission is dominated by non-adiabatic switch-on, the
acceleration no longer plays a crucial role''\cite{HuRoura}.

\section{Conclusion}

A quantum system that undergoes linear uniform acceleration
in vacuum perceives a thermal bath at the Unruh temperature $a/2\pi$\cite{Unruh}.
If this system is bound to interact with the surrounding quanta in a finite size cavity, one expects the picture to differ
from the original freely propagating one.
Indeed, if the cavity is a closed rigid box, the configuration of the vacuum itself is different, because of the Casimir effect\cite{Casimir}.
Moreover, entering the cavity, or equivalently switching on abruptly the interaction,
is a non-adiabatic process that alters the transition amplitudes drastically.
In order to treat this problem, one can choose to consider ``smooth'' boundary conditions
by switching the interaction progressively on and off along the trajectory
with the help of a gate-like function $f(\t)$
that enters in the Hamiltonian expression, \reff{Hint}.
This was our choice along the present paper.
This way, the definition of the vacuum and the particle states stay unmodified with respect to the free case,
one can use the interacting picture and focus solely on the switching effect.
The slope of $f(\t)$ is an estimate of the switching non-adiabaticity
and the transition amplitudes vary according to the interplay between the detector gap $E$ and
the Doppler-shifted quanta energy $k(1\pm\cos\theta)e^{\pm a\t}/2$.

The main results of this paper are Eqs.(\ref{ratiooffaxissmooth})
and (\ref{ratiosemiinfonaxis}-\ref{TheRelation}).
They are obtained by introducing a simple switching function (\ref{superf})
that enables analytical expressions throughout.
The first equation gives the emission/absorption ratio for any off-axis quanta,
in a cavity that a uniformly accelerated detector enters at $\t_i$ and leaves at $\t_f$.
When these times are repelled at infinities, one recovers the Unruh result,
that is, a $\vec{k}-$independent thermal ratio.
Otherwise, the ratio depends on the quanta momentum,
and the pseudo-temperature we obtain is {\it always greater} than the Unruh temperature.
It is hardly acceptable to keep on calling this behavior ``Unruh effect''
since neither thermality is achieved nor uniformity of the coupling is respected[27].

The same conclusion applies to the result (\ref{ratiosemiinfonaxis})
which concerns co-propagating quanta in semi-infinite cavities.
Contrary to the sole value $r^{UA}_{\vec{k}}=a/2\pi E$
which has been proposed in the literature\cite{Scully1,Scully2,Scully3},
we find again that the situation is only pseudo-thermal:
Provided the detector enters the cavity late enough (\ie with a sufficiently large velocity),
the pseudo-temperature is increased as much as one wishes beyond the Unruh temperature, see Fig.(\ref{Fig4}).

Finally, \reff{TheRelation} shows that
the pseudo-temperature is formally independent of the acceleration
since its expression applies as well to inertial trajectories
in some cavities.

Moreover, one can learn two technical lessons from the present work
that may be applied to the experimental study of the Unruh effect\cite{Exp}.
First, abrupt boundary conditions are a dangerous choice since they lead to artefact amplitudes
that provide extreme values of the emission/absorption ratio (see Appendix A);
smooth switchings are more advisable although their structure has to be well-defined.
In the picture proposed in \cite{ChenTajima}, it means that
one should have the undergone acceleration precisely known
around the selected nodes of the electro-magnetic pattern
in order to foresee the emitted radiation.
Secondly, one should treat with extra care co-propagating quanta since
they experience the most extreme Doppler effects (see Appendices B2 and C2).
For instance, one could avoid such problems by mode-blocking only transverse photons\cite{Habs}.

${\bf{Acknowledgements}}$
\newline
The author thanks R.Parentani for useful advice, D.Campo for endless fruitful comments
and A.Schobert for careful rereading, as well as an anonymous referee for enlightening questions.
The author is a Feinberg Graduate School fellow.

\appendix
\section{Gate and smooth gate switching functions in the inertial case}

\subsection{Gate function}

Mimicking a cavity of length $L$ for a detector travelling at speed $v$ means that $f(\t)$ has to be
extremely small outside $[\t_i,\t_i+L/\gamma v]$ and nearly $1$ inside.

If we suppose that the interaction is strictly uniform inside the cavity and totally vanishes outside,
the interaction is described by the pure gate ($G$) function
$f_{G}(\t)=\Theta(\t-\t_i)\Theta(\t_f-\t)$, which straightforwardly provides
\ba\label{fourierinert}
{}^{G}f_\om = \frac{\sin\left(\om\Delta\t/2\right)}{\pi\om }\: e^{i \om (\t_i+\t_f)/2}
\ea
and
\ba\label{rinertunif}
{}^{G}r^{I}_{\vec{k}} &=&
\left|\frac{\sinc\left[(E+k
\xi(\theta))\Delta\t/2\right]}{\sinc\left[(E-k
\xi(\theta))\Delta\t/2\right]}\right|^2\ ,
\ea
where $\sinc(x)=\sin x/x$ is the cardinal sine and $\Delta\t=\t_f-\t_i=L/\gamma v$.
Therefore, for detectors
such that their gap satisfy $(\pm E+ k \xi(\theta))\Delta\t=2\pi
n, \; n\in \mathbb{N}^*$, the ratio ${}^{G}r^{I}_{\vec{k}}$
is zero (respectively infinite). This seems to mean that, given $E$ and
according to the values of $\vec{k},L$ and $v$, an ensemble of
detectors can be either altogether in the lower state or excited
when leaving the cavity, at least at the first order in the
perturbative expansion.
These resonant energies do not exist any longer when integrating over $\theta$
since ${}^{G}r^{I}_{k}$ is always finite:
\ba\label{rinertkgate}
{}^{G}r^{I}_{k}=\frac{h\left[(E + k e^{\nu})\Delta\t\right]-h\left[(E + k e^{-\nu})\Delta\t\right]}
{h\left[(E - k e^{-\nu})\Delta\t\right]-h\left[(E - k e^{\nu})\Delta\t\right]}
\ea
with $h(x)=(\cos(x)-1)/x + \Si(x)$, where $\Si(x)=\int_{0}^{x} \! \mbox{d} y \; \frac{\sin(y)}{y}$
is the Sine Integral.
As $h(x)$ is a monotonic increasing function, neither the denominator nor the numerator of
\reff{rinertkgate} can vanish.
However, ${}^{G}r^{I}_{k}$ can still take values above $1$.
Finally, the integral expression of ${}^{G}r^{I}$ is diverging in the
UV limit\cite{footnotespecialcase}.
From this we learn that a perfect gate describing a cavity with abrupt boundaries is a
pathological choice that leads to resonant effects, even in the simplest case of an inertial trajectory.
{\it A fortiori}, such a choice should be avoided for more complicated world-lines.

\subsection{Smooth gate function}

\begin{figure}[h!]
\hbox to\linewidth{\hss
\resizebox{7.6cm}{4.7cm}{\includegraphics{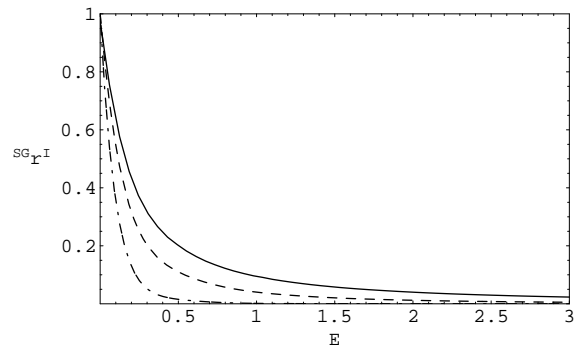}}
\hss}
\caption{The ratio ${}^{SG}r^{I}$ as a function of $E$
when $\Delta\t=10$
and for different values of the switching lapse $\delta=0.01$ (plain), $\delta=0.1$ (dashed) and $\delta=1$ (dashed dot).
The ratio is always below $1$ and decreases to $0$ as expected.
Moreover, the shorter the switching time, the higher ${}^{SG}r^{I}$.
This is the signature of transient effects.
\label{Fig3}}
\end{figure}

In order to avoid such effects one can describe the cavity by smoother boundaries.
We show that, in this case too, a great care has to be taken.
To this end, let us generalize the gate function and
introduce a smooth gate ($SG$) and its Fourier components:
 \ba
   f_{SG}(\t)&=&\half\left(\tanh\left[\frac{\t-\t_i}{\delta}\right]-\tanh\left[\frac{\t-\t_f}{\delta}\right]\right) \\
   {}^{SG}f_\om&=&\frac{\delta}{2}\frac{\sin\left(\om\Delta\t/2\right)}{\sinh\left(\pi\om\delta/2\right)} \: e^{i\om(\t_f+\t_i)/2} \ .
   \label{Fouriersmoothgate}
 \ea
$\delta$ is the typical lapse during which the interaction is switched on (around $\t_i$) and off (around $\t_f$);
between $\t_i$ and $\t_f$, $f(\t)\simeq 1$, outside this interval it is exponentially small;
when $\delta\to 0$, one recovers the gate function (and \reff{Fouriersmoothgate} tends to \reff{fourierinert}).
Using this function, the emission/absorption ratio is given by
\ba\label{rinertvecksinh}
{}^{SG}r^{I}_{\vec{k}}
= {}^{G}r^{I}_{\vec{k}} \times
\left| \frac{\sinhc\left[\pi(E-k \xi(\theta))\delta/2\right]}{\sinhc\left[\pi(E+k \xi(\theta))\delta/2\right]}\right|^2
\ea
where $\sinhc(x)=\sinh x/x$ is the cardinal hyperbolic sine.
One clearly sees that ${}^{SG}r^{I}_{\vec{k}}$ also exhibits vanishing and infinite values
since the periodic zeros and poles of ${}^{G}r^{I}_{\vec{k}}$ are not compensated by the non-periodic multiplying factor.
However, the entirely integrated ratio ${}^{SG}r^{I}$ is finite and always less than one, as shown in Fig.\ref{Fig3}.
With this respect, one realizes that even a smooth interaction can lead to suspicious expressions
for some specific wave-vectors, but that resonant effects may disappear after integration over $\theta$ and $k$.
This example also tells us that, for {\it fixed} cavity and quanta properties (same $L$ and $\xi$),
the corresponding ratios differ significantly while varying the switching $\delta$.

The conclusion of this appendix is that one should be extremely careful when modelling a cavity:
one would better avoid abrupt boundary conditions, of course, but one should be aware that even smooth
switching functions can generate pathological ratios.

\section{The Unruh amplitudes}

In this appendix, we explicit how the amplitudes are obtained
when the detector is uniformly accelerated and uniformly coupled to the field,
\ie $f(\t)\equiv 1$.

\subsection{Off-axis quanta}

According to \reff{Aunifacccpmgeneral},
the emission and absorption amplitudes are proportional to the following integrals:
\ba
I_\pm &=&\Ie{\t} e^{i (\pm E \t + k\frac{1-\cos\theta}{2a} e^{a\t} - k\frac{1+\cos\theta}{2a} e^{-a\t} )} \label{eqapp2}
\ea
One can check that this expression is convergent by making the change of variables $t=e^{a\t}$.
One gets $\Io{t} t^{-1\pm iE/a} e^{ik_-t-ik_+/t}$.
The behaviors at both limits are of the same nature (it can be seen by changing $t$ for $t'=1/t$).
The oscillatory exponential term ensures convergence for $t\to\infty$ (\eg as for a Cosine Integral).

When using $x=a\t-\al$ and $k_\pm\equiv k(1\pm\cos\theta)/2a$ one gets
\ba
I_\pm &=& \frac{e^{\pm i E\al/a}}{a} \int_{-\infty-\al}^{+\infty-\al}\! \mbox{d} x \; \nn
&& \times \: e^{i (\pm E x /a + k_-e^\al e^x - k_+e^{-\al}e^{-x})}\nn
&=& \frac{e^{\mp \pi E/2a}}{a}  {\left(\frac{1+\cos\theta}{1-\cos\theta}\right)}^{\pm iE/2a} \nn
&& \times \int_{-\infty-i\pi/2}^{+\infty-i\pi/2}\! \mbox{d} y \; e^{-ze^y-ze^{-y}\pm iE y/a} \nonumber
\ea
where the second expression is obtained by choosing the parameter $\al$ such that
$\al = i\pi/2 + \ln\sqrt{(1+\cos\theta)/(1-\cos\theta)}$, and noting $z=\sqrt{k_+ k_-}=k(\sin\theta)/2a$,

By contour integration, one can show that the last integral can be dragged upon the real axis
since it possesses no pole and the boundary terms vanish.
Therefore, the amplitudes are mathematically well-defined and,
since $2K_\nu(2z)=\Ie{y}e^{-ze^{-y}-ze^{y}+\nu y}, \; \re(z)>0$, one gets \reff{Aunifaccpmoffaxis}.

\subsection{Aligned quanta}

According to \reff{Aunifacccpmgeneral},
the relevant integrals for the on-axis quanta differ
if we consider the creation of a co-propagating ($p.$) quanta or that of counter-propagating ($c.p.$) ones:
\ba
I_\pm^{p.} &=& \Ie{\t} e^{i (\pm E \t - k \, e^{-a\t}/a)} \nn
I_\pm^{c.p.} &=& \Ie{\t} e^{i (\pm E \t + \frac{k}{a} \, e^{a\t})} \ .\nonumber
\ea
Let us focus on the co-propagating emission amplitude expression:
\ba
I_+^{p.} 
&=& \inv{a} \left(\frac{k}{a}\right)^{iE/a} \: \Io{x} x^{-1-iE/a} \: e^{-i x} \nn
&=& \inv{a} \left(\frac{k}{a}\right)^{iE/a} \! e^{-\pi E /2 a} \:
\int_{0}^{i \infty}\! \mbox{d} y \; y^{-1-iE/a} \: e^{-y}\nonumber
\ea
where we used $x=ke^{-a\t}/a$ to obtain the second integral and
$y=e^{i\pi/2} x$ to obtain the third one.
Contrary to the off-axis case, one can see from the second formulation above that
this integral diverges in the $x\to 0$ limit.
A way to cure this divergency is to replace $-iE/a$ by $-iE/a+\delta$,
where $\delta$ is any small positive constant.
Let us see in details how it should be done.
The third integral is the limit of
\ba
\int_{i\e}^{i R}\! \mbox{d} y \; y^{-1-iE/a} \: e^{-y}
&=& i R^{-iE/a} \: \int_{0}^{\pi/2}\! \mbox{d} \theta \; e^{E\theta/a} \: e^{-Re^{i\theta}} \nn
&-& i \e^{-iE/a} \: \int_{0}^{\pi/2}\! \mbox{d} \theta \; e^{E\theta/a} \: e^{-\e e^{i\theta}} \nn
&+ & \int_{\e}^{R}\! \mbox{d} y \; y^{-1-iE/a} \: e^{-y} \ ,\nonumber
\ea
for $\e\to 0$ and $R\to+\infty$, where the right-hand side has been obtained by
contour integration.
When extending analytically the properties of the energy into the complex plane,
and requiring that $E=E+ i a \delta, \; 0<\delta\ll 1$,
one obtains the following behaviors for the right-hand side of the preceding expression.
The first term behaves as $\frac{R^\delta}{R+ E/a} e^{\pi E/2a}\to 0$ when $R\to\infty$;
as we pointed out, the $\delta$ term is not mandatory for the large modulus convergence.
The second term behaves as $\e^\delta \frac{a}{E} (e^{\pi E/2a}-1)\to 0$ when $\e\to 0$,
thanks to the $\e^\delta$ decaying term.
Finally, the last term tends to $\Gamma(-iE/a + \delta)$ since the Gamma function is defined by
$\Gamma(x)=\Io{t}t^{-1+x}e^{-t}, \, \re(x)>0$.

The integrals involved in the other amplitudes are
\ba
I_+^{c.p.} &=& \inv{a} \left(\frac{k}{a}\right)^{-iE/a} \: e^{-\pi E /2 a} \:
\int_{0}^{-i \infty}\! \mbox{d} y \; y^{-1+iE/a} \: e^{-y} \nn
I_-^{p.} &=& \inv{a} \left(\frac{k}{a}\right)^{-iE/a} \: e^{+\pi E /2 a} \:
\int_{0}^{i \infty}\! \mbox{d} y \; y^{-1+iE/a} \: e^{-y} \nn
I_-^{c.p.} &=& \inv{a} \left(\frac{k}{a}\right)^{iE/a} \: e^{+\pi E /2 a} \:
\int_{0}^{-i \infty}\! \mbox{d} y \; y^{-1-iE/a} \: e^{-y} \ ,\nonumber
\ea
and suffer from the same divergency as $I_+^{p.}$.
By requiring that $E=E\pm i a \delta, \; 0<\delta\ll 1$ respectively for co- and counter-propagating quanta
for the emission amplitudes (and inversely for the absorption amplitudes), one can have $\e\to 0$ and $R\to+\infty$
and write down the expression of all amplitudes
\ba
{}^{p.}A_{\pm,\vec{k}}^{UA}&=& \frac{- i g}{\sqrt{2{(2\pi)}^3k}} \;\label{Aunifaccccopropdelta}
\inv{a} \left(\frac{k}{a}\right)^{\pm iE/a-\delta} \: \\
&& \times e^{\mp\pi E /2 a-i\pi\delta/2} \: \Gamma(\mp iE/a+ \delta)  \nn
{}^{c.p.}A_{\pm,\vec{k}}^{UA}&=& \frac{- i g}{\sqrt{2{(2\pi)}^3k}} \;\label{Aunifacccountpropdelta}
\inv{a} \left(\frac{k}{a}\right)^{\mp iE/a-\delta} \: \\
&& \times e^{\mp\pi E /2 a+i\pi\delta/2} \: \Gamma(\pm iE/a+ \delta) \ .\nonumber
\ea
Given that\cite{Abra}
\ba
\Gamma(i\al+ \delta)^*=\Gamma(-i\al+ \delta) \ ,\nonumber
\ea
we find the celebrated Unruh result :
\ba\label{r.unif.acc.aligned}
r^{UA}_{\vec{k}}&=&\left|\frac{{}^{p.}A_{+,\vec{k}}^{UA}}{{}^{p.}A_{-,\vec{k}}^{UA}}\right|^2
= \left|\frac{{}^{c.p.}A_{+,\vec{k}}^{UA}}{{}^{c.p.}A_{-,\vec{k}}^{UA}}\right|^2\\
&=& \left|\frac{{}^{p.}A_{+,\vec{k}}^{UA}+{}^{c.p.}A_{+,\vec{k}}^{UA}}
{{}^{p.}A_{-,\vec{k}}^{UA}+{}^{c.p.}A_{-,\vec{k}}^{UA}}\right|^2
= e^{-2\pi E /a } \nonumber \ ,
\ea
independently of $\delta$, although the $\delta\to 0$ limit has not been taken.

Note that conventionally, the limit $\delta\to 0$ is taken and the amplitudes
involve a mere $\Gamma(\pm iE/a)$ term in the literature.
However one should remember that, if this expression mathematically exists
for example as the ratio $\Gamma(1\pm iE/a)/(\pm iE/a)$, or as a series representation,
it does not as the integral representation $\Gamma(z)=\Io{t}t^{-1+z}e^{-t}$.

Note also that the $p.$ (resp. $c.p.$) above amplitudes are
the $\cos\theta\to 1$ (resp. $-1$) limit of the off-axis amplitudes \reff{Aunifaccpmoffaxis},
when using the small argument behavior of the Bessel functions\cite{Abra}
$2\: K_\nu(2z)\simeq z^\nu \Gamma(-\nu)+z^{-\nu} \Gamma(\nu)$ and the same $\delta$ prescription.

\section{The Modified Amplitudes}

In this appendix, we explicit how the amplitudes are obtained
when the detector is uniformly accelerated and coupled to the field
via the function \reff{superf}.

\subsection{Off-axis quanta}

In the case of off-axis quanta, the method is equivalent to the one
we expounded in the uniform coupling regime.
The integrals to be expressed are
\ba\label{amplapp3}
I_\pm &=&\Ie{\t} e^{-\eta_1 e^{-a\t}-\eta_2 e^{a\t}} \\
&& \times e^{i (\pm E \t + k(1-\cos\theta) e^{a\t}/2a - k(1+\cos\theta) e^{-a\t}/2a )} \nn
&=& \frac{e^{\pm i E\al/a}}{a} \int_{-\infty-\al}^{+\infty-\al}\! \mbox{d} x \;\nn
&& e^{\pm i E x /a -(\eta_2 - i k_-)e^\al e^x - (\eta_1 + i k_+)e^{-\al} e^{-x})} \ ,\label{C2}
\ea
where we used $x=a\t-\al$.
By choosing the parameter $\al$ such that
$e^{2\al}=(\eta_1+ik_+)/(\eta_2-ik_-)$, \ie
$\al = \ln\sqrt{({\eta_1}^2+{k_+}^2)/({\eta_2}^2+{k_-}^2)} + \frac{i}{2} \arctan(k_+/\eta_1)+ \frac{i}{2} \arctan(k_-/\eta_2)$,
and noting $z=({\eta_1}^2+{k_+}^2)^{1/4}({\eta_2}^2+{k_-}^2)^{1/4} e^{\frac{i}{2}(\arctan(k_+/\eta_1)-\arctan(k_-/\eta_2))}$,
one obtains
\ba
I_\pm &=& \inv{a} {\left(\frac{{\eta_1}^2 + {k_+}^2}{{\eta_2}^2 + {k_-}^2}\right)}^{\pm iE/4a}
\int_{-\infty-\al}^{+\infty-\al}\! \mbox{d} y \; e^{-ze^y-ze^{-y}\pm iE y/a}
 \nn
&& \; \; \times \;  e^{\mp \frac{E}{2a}(\arctan(k_+/\eta_1)+\arctan(k_-/\eta_2))}\ .\nonumber
\ea
Once again, the remaining integral can be dragged on the real axis and therefore
one obtains \reff{Aunifaccsmoothoffaxis}.
Interesting is to remark that $\eta_1$ and $\eta_2$ of \reff{superf} actually also
define the analytical content of $k_\pm$, as one can see in \reff{C2}:
$k_+\to k_+-i\eta_1\:,\:k_-\to k_-+i\eta_2$.
On the top of the trivial crossing symmetry relation $A_-(-E)=A_+(E)$,
one has the additional
${}^{cav}A_{\pm,\vec{k}}^{UA}(k_+,\eta_1;k_-,\eta_2)
=-[{}^{cav}A_{\pm,\vec{k}}^{UA}(k_-,\eta_2;k_+,\eta_1)]^*$.
This ensures that the pair-creation process is invariant
under time-reversal, defined as $\theta\to\pi-\theta$
and $\t_{i,f}\to-\t_{f,i}$.

\subsection{Aligned quanta}

Before computing the corresponding expressions, we would like to focus on
the fact that the rapid decay of the switching function
\reff{superf} ensures the convergence of all expressions.
Therefore, contrary to the everlasting interaction case, the expressions
will not involve ill-defined Gamma functions but well-behaved Bessel ones.
Nevertheless, taking the limit $\eta_1\to 0$ or $\eta_2\to 0$ will inevitably
lead to incoherences, as we shall show.

Let us focus on the co-propagating quanta. The relevant integral is
\ba\label{integrale}
\Ie{\t}e^{-\eta_1 e^{-a\t} -\eta_2e^{a\t}} \: e^{i(\pm E\t -ke^{-a\t}/a)}
\ea
which is simply the expression \reff{amplapp3} when $\theta=0$.
We can blindly apply the same method as before to all the amplitudes, noting the
corresponding results by an overall $\tilde{ }$ :
\bwt
\ba
{}^{p.}_{}\tilde A_{\pm,\vec{k}}^{UA} &=& \frac{- 2 i g}{a\sqrt{2{(2\pi)}^3k}} \:
(\eta_1^2+k^2/a^2)^{\pm iE/4a}\eta_2^{\mp iE/2a} e^{\mp \frac{E}{2a}\arctan(k/a\eta_1)} \nn
\label{Aunifaccsmoothpmpr}
&& \times \; \; K_{\pm iE/a}[2(\eta_1^2+k^2/a^2)^{1/4}\eta_2^{1/2} e^{\frac{i}{2}\arctan(k/a\eta_1)}] 
\ea
\ewt
Let us compute the emission/absorption ratio from the previous expression.
Since $K_\nu(z)=K_{-\nu}(z)$ for any $(\nu,z)$, the Bessel functions for emission and absorption
are equal and the ratio is given by
\ba
{}^{p.}_{}\tilde r^{UA}_{\vec{k}}&=&
e^{-\frac{2E}{a}\arctan(k/a\eta_1)} \nn 
\ea
First, note that the co-propagating ratio does not seem to depend on the end of the interaction.
This comes from the fact that the corresponding Doppler factor pathologically vanishes.
Secondly, this ratio tends to $e^{-\pi E /a }$ when the switching on times is sent to $-\infty$.
This is the square root of the Unruh result one should recover.
However, we know already from the previous appendix that the co-propagating integral representation
is ill-defined if the switching off time is sent to infinity.
Therefore, although ${}^{p.}r$ does not seem to depend on $\eta_2$, one can not take the limit $\eta_2\to 0$ safely.
This can be seen by remarking that the argument of the Bessel functions tends to zero in this case.
If one wishes to recover the Unruh limit by sending the time limits of the interaction to infinity,
one has to remember that an imaginary part has to be given to the energy to make the integrals convergent.
The $\eta_2\to 0$ limit causes the argument of the Bessel functions to be small.
Therefore, thanks to the relation
$2K_\nu(2z) \simeq z^\nu \ \Gamma(-\nu) + z^{-\nu} \ \Gamma(\nu)$ for small $z$, one finds
\bwt
\ba
{}^{p.}_{}A_{\pm,\vec{k}}^{UA}
&\simeq&
\frac{- i g}{a\sqrt{2{(2\pi)}^3k}} \:
(\eta_1^2+k^2/a^2)^{\pm iE/4a-\delta/4}\eta_2^{\mp iE/2a+\delta/2} e^{\mp \frac{E}{2a}\arctan(k/a\eta_1)-i\frac{\delta}{2}\arctan(k/a\eta_1)} \nn
&& \times \left[
(\eta_1^2+k^2/a^2)^{\pm iE/4a-\delta/4} \eta_2^{\pm iE/2a-\delta/2} e^{\mp \frac{E}{2a}\arctan(k/a\eta_1)-i\frac{\delta}{2}\arctan(k/a\eta_1)} \:
\: \Gamma(\mp iE/a+\delta) \right. \nn
&& \left. \;\;\; +
(\eta_1^2+k^2/a^2)^{\mp iE/4a+\delta/4} \eta_2^{\mp iE/2a+\delta/2} e^{\pm \frac{E}{2a}\arctan(k/a\eta_1)+i\frac{\delta}{2}\arctan(k/a\eta_1)} \:
\: \Gamma(\pm iE/a-\delta) \right] \nn
&\simeq&
\frac{- i g}{a\sqrt{2{(2\pi)}^3k}} \nn
&& \times \left[
(\eta_1^2+k^2/a^2)^{\pm iE/2a-\delta/2} e^{\mp \frac{E}{a}\arctan(k/a\eta_1)-i\delta\arctan(k/a\eta_1)} \:
\: \Gamma(\mp iE/a+\delta) \right. \nn
&& \left. \;\;\; +
\eta_2^{\mp iE/a+\delta} \: \Gamma(\pm iE/a-\delta) \right] \nn
&\stackrel{\eta_2\to 0}{\longrightarrow}& \frac{- i g}{a\sqrt{2{(2\pi)}^3k}} \nn
&\times& \left[
(\eta_1^2+k^2/a^2)^{\pm iE/2a-\delta/2} e^{\mp \frac{E}{a}\arctan(k/a\eta_1)-i\delta\arctan(k/a\eta_1)} \:
\: \Gamma(\mp iE/a+\delta) \right] \ .\label{Aunifaccccopropdeltasmooth}
\ea
\ewt
Thus, for any $\eta_1$ and for $\eta_2=0$, the ratio is given by
\ba\label{ratiocoprosmooth}
{}^{p.}_{}r^{UA}_{\vec{k}}
=  e^{\disp -\frac{4 E}{a}\arctan(k/a\eta_1)}\ ,
\ea
independently of the $\delta$.
One easily can check that the $\eta_1\to 0$ limit for \reff{Aunifaccccopropdeltasmooth}
provides the Unruh amplitude \reff{Aunifaccccopropdelta}
as the ratio \reff{ratiocoprosmooth} tends to $e^{-\frac{2\pi E}{a}}$.
Obviously, the same results are obtained for the counter-propagating amplitudes,
where the roles of $\eta_1$ and $\eta_2$ are exchanged.


\begin{thebibliography}{99}
\bibitem{footnotef}
Actually, the maximum value is $e^{-2\sqrt{\eta_1\eta_2}}$
and is obtained at $(\t_i+\t_f)/2=\ln(\sqrt{\eta_1/\eta_2})/a$.
Therefore, the plateau shape,
its value one and its limits $\t_i$ and $\t_f$ are a good approximation
only in the regime where $\eta_1\eta_2$ is sufficiently small,
\ie when the duration of the interaction $\t_f-\t_i$ is sufficiently large with respect to that of the switching $1/a$.
This is the regime we consider from now on.
\bibitem{footnotevisser}
As for the quarrel about whether this kind of departure from the
original Unruh picture can still be called Unruh effect or is just
a mere (but interesting) quantum field phenomenon (since it
involves non-adiabatic physics), we follow Visser\cite{Visser} and
let the debate to linguistic experts.
\bibitem{footnoteWKB}
In fact, one could also bypass the necessity of recovering analytical expressions
by inspecting the WKB expressions obtained from \reff{Aunifacccpmgeneral}\cite{primer}.
However, as all approximations, the WKB one is efficient only for a certain range of parameters
and we wish our treatment to depend as less as possible on the wave-vector $k$.
\end{thebibliography}
\end{document}